\documentclass[aip,jap,reprint]{revtex4-1}

\usepackage{graphicx}
\usepackage{epstopdf} 
\usepackage{amsmath}
\usepackage[bf]{subfigure}

\begin{document}

\title{Effect of disorder studied with ferromagnetic resonance for arrays of tangentially magnetized sub-micron Permalloy discs fabricated by nanosphere lithography}

\author{N. Ross}
\email{rossn2282@gmail.com}
\author{M. Kostylev}
\email{kostylev@cyllene.uwa.edu.au}
\author{R. L. Stamps}
\affiliation{School of Physics, University of Western Australia, Crawley, WA, Australia}

\date{\today}
 
\begin{abstract}
Tangentially magnetized trigonal arrays of sub-micron Permalloy discs are characterized with ferromagnetic resonance to determine the possible contributions to frequency and linewidth from array disorder. Each array is fabricated by a water-surface self-assembly lithographic technique, and consists of a large trigonal array of 700 nm diameter magnetic discs. Each array is characterized by a different degree of ordering. Two modes are present in the ferromagnetic resonance spectra: a large amplitude, `fundamental' mode and a lower amplitude mode at higher field. Angular dependence of the resonance field in a very well ordered array is found to be negligible for both modes. The relationship between resonance frequency and applied magnetic field is found to be uncorrelated with array disorder. Linewidth is found to increase with increasing array disorder.

\end{abstract}

 
\maketitle

\section{Introduction}

There is currently an intense interest in the magnetic properties of nanoscale and sub-micron discs of low aspect ratio. Such discs have potential applications in data storage,\cite{TerrisandThomson2005, Thomsonetal2006,Shawetal2007, Hellwigetal2007, Shawetal2008} spintronics,\cite{Kiselevetal2003, Deacetal2008} and medicine,\cite{Zabowetal2008, Kimetal2009} and are otherwise viewed as simple model systems by which the properties of more exotic nanostructures can be understood. In many fundamental studies and in data storage and spintronics applications in particular, the dynamic magnetic properties of the disc structures are particularly important. Where arrays of discs are densely packed, dipole coupling between discs can have a significant impact on dynamic behaviour.  There have been a variety of experimental and theoretical studies of the dynamic magnetic properties of arrays of dipole-coupled magnetic discs, with recent publications concerning resonance frequency position with respect to applied magnetic field,  \cite{Kakazeietal2006, Kostylevetal2008IEEE, Nevirkovetsetal2008, Tsaietal2009} mode structure in the vortex state,\cite{ShibataandOtani2004,Galkinetal2006} and mode structure for in-plane \cite{Jorzicketal1999, GuslienkoandSlavin2000, Jungetalexp2002, Jungetaltheory2002, PolitiandPini2002, Rivkinetal2004,  GubbiottietalJAP2006, Giovanninietal2007, RivkinetalPRB2007, RivkinetalJMMM2007, Nevirkovetsetal2008} and out-of-plane  \cite{Kakazeietal2004, Mewesetal2006, Kostylevetal2008IEEE} applied magnetic fields. Recognition of the importance of spin-wave damping for many applications has led to a number of studies into ferromagnetic resonance (FMR) linewidth broadening in such systems, \cite{Schneideretal2007, Rivkinetal2009, Shawetal2009} and recently in a study of microwave-assisted switching of magnetic array nano-elements. \cite{Nembachetal2009} To date, however, an experimental study of the effects of array packing on damping of spin wave modes observable with FMR has been lacking.

One approach to such a study is to measure FMR linewidth damping in  arrays that were patterned by a `top-down' method like Focussed Ion Beam lithography in such a way that they had identical disc geometries but varied pitch (inter-disc spacing). Such fabrication techniques are time-intensive and therefore often only suitable for the production of small arrays ($<1~\mathrm{mm^2}$). The FMR responses of small arrays are correspondingly small. Such small signals are often beyond the sensitivity of conventional FMR techniques, so that the previous experimental studies of damping in arrays of nano-discs utilized time-resolved Kerr microscopy \cite{Schneideretal2007, Shawetal2009} and `meander-line' FMR \cite{Rivkinetal2009} for the detection of spin wave modes . In this study, nanosphere lithography has been used to facilitate the production of large-area arrays, thus allowing measuring using conventional Vector Network Analyzer FMR (VNA-FMR).

`Nanosphere lithography' refers broadly to techniques in which the tendency of colloidal particles of sub-micron or nanoscale size to self-assemble on hydrophilic surfaces is exploited to produce a lithographic mask; the relevant literature is replete with various examples of such techniques. \cite{DeckmanandDunsmuir1982, HulteenandVanDuyne1995, Hulteenetal1999,Jensenetal1999, Haginoyaetal1997, Burmeisteretal1997, Ormondeetal2004, Rybczynskietal2003, Weekesetal2004, Weekesetal2007, Lietal2008, Zhangetal2008} The most pronounced limitation of such techniques is the difficulty of achieving long-range ordering in the colloidal mask.

In this study, this tendency towards partial disorder was used to fabricate a series of array samples with varied degrees of array ordering. A nanosphere lithographic technique involving assembly of colloidal particles into a monolayer on a water surface \cite{Weekesetal2007} was used to produce arrays of sub-micron magnetic discs with deliberately varied degrees of array order but the same local disc-to-disc spacing and disc geometry. The FMR mode structures and linewidths of this series of array samples were measured by VNA-FMR.

\section{Experiment}

\begin{figure}[htbp]
\begin{center}
\includegraphics[width=8.5cm]{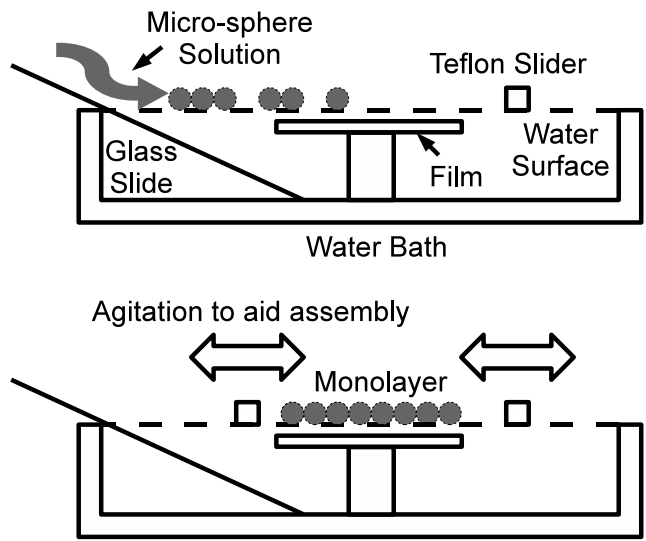}
\caption{\label{watersurfaceassembly} Schematic cross-section representation of the deposition of the lithographic mask, followed by agitation of the water surface with laterally placed teflon sliders to aid assembly of trigonal monolayers.}
\end{center}

\end{figure}

Four sub-micron magnetic disc array samples with different degrees of ordering were fabricated by a technique derived from that of Weekes \emph{et al.} \cite{Weekesetal2007} Three continuous $\mathrm{Ni_{81} Fe_{19}}$ (Permalloy) films, thickness $L = 27 \pm 3$ nm, were radio-frequency magnetron sputtered on silicon substrates with 40 nm Ta seed layers: these films are denoted f1, f2, and f3 in this study. Sputtered thickness was verified by the measurement of step edges of independently sputtered films using white-light optical profilometry. Each of these films was placed on a raised platform in a narrow bath filled with 18 $M \Omega \cdot \mathrm{cm}$ water, so that the Permalloy face of the film was under the water surface but as close to it as possible. Sodium Dodecyl Sulphate (SDS) was added to this bath to a concentration of 150 $\mu \mathrm{g~L^{-1}}$: it has been demonstrated in the literature that SDS can aid the assembly of reagent microspheres on a water surface. \cite{Lietal2008, Zhangetal2008} 780 nm diameter carboxylate-terminated polystyrene reagent microspheres obtained from Duke Scientific were mixed with reagent-grade ethanol into a 1:2 microsphere solution : ethanol mixture by volume. This mixture was introduced to the water surface via a hydrophilic glass slide previously cleaned in SDS. Teflon sliders placed laterally across the water bath were used to gently agitate the water surface, aiding in the self-assembly of well ordered trigonal monolayers of microspheres. These sliders were used to position the monolayers above the film, and the water drained quickly from the bath in order to transfer the monolayer to the film surface. Varying the degrees of agitation and of monolayer compression during the transfer stage allowed control over the degree of ordering of the resulting lithographic mask. The assembly of the lithographic mask on the water surface is depicted diagrammatically in Figure \ref{watersurfaceassembly}.

Once masked with the trigonal array of microspheres, several $5 \times 8~\mathrm{mm^2}$ sections were cut from each of the parent films f1, f2, and f3: one such section from each parent film was left masked but was not patterned; the other sections were patterned. In this study, un-patterned sections of the parent film are denoted f1c, f2c, and f3c. Four films with different degrees of mask ordering were patterned: in this study they are denoted f1a, f2a, f3a, and f3b to make clear the identity of the film from which each was patterned. Each film was patterned by placing it in the target position of a radio-frequency sputtering chamber which was evacuated to a base pressure of $1.0 \times 10^{-6}$ Torr. Samples were reactively ion etched in an oxygen atmosphere of 100 mTorr at a power density of $0.5 ~\mathrm{W~cm^{-2}}$ in order to reduce the diameter of the mask units. Following etching, each sample was argon milled at  a pressure of 25 mTorr and power density of $4.25~\mathrm{W~cm^{-2}}$ to remove Permalloy not masked by microsphere material.

 \begin{figure}[htbp]
\begin{center}
\includegraphics[width=8.5cm]{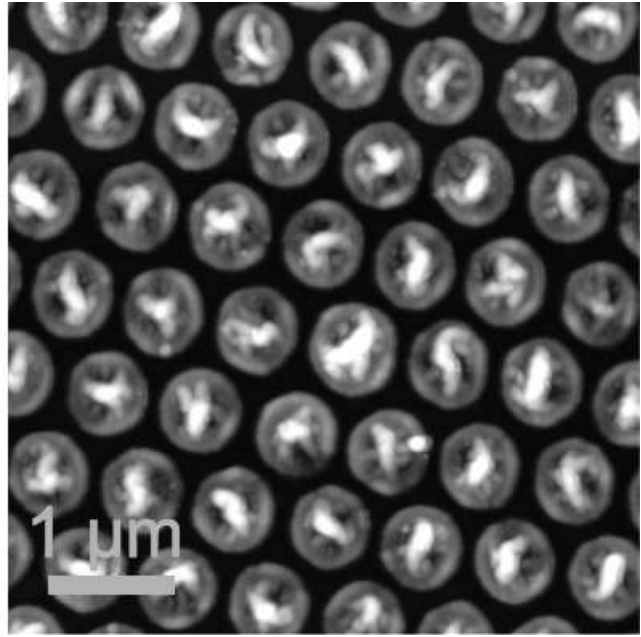}
\caption{\label{s144bAFM} AFM image representative of the microscale character of the disc arrays; the high contrast regions in the centre of the discs indicate that some of the polystyrene caps remain. This particular image is a 5 $\times$ 5 $\mu$m image of sample f3b.}
\end{center}

\end{figure}

 \begin{figure}[htbp]
\begin{center}
\includegraphics[width=8.5cm]{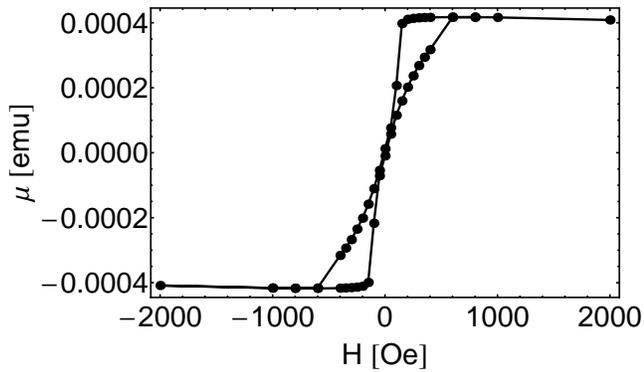}
\caption{\label{s144bhys} SQuID-measured hysteresis loop of f3b, typical of the loops recorded for all the patterned films; the fit lines in this figure are linear interpolations that serve as a guide for the eye. The hysteresis loop shows the two closed loops characteristic of vortex nucleation, movement, and annihilation. The narrowness of the `pinching off' of the two loops provides confirmation for the whole sample of smooth dot boundaries seen in Figure \ref{s144bAFM}.}
\end{center}

\end{figure}
 
 Characterization of disc geometry was achieved with tapping-mode Atomic Force Microscopy (AFM): an example AFM image is shown in Figure \ref{s144bAFM}. With the scan parameters used, the AFM could be used to resolve structures down to 20 nm. The high structures in the centre of the discs in the AFM image are the remains of polystyrene caps. The discs in all of the arrays have very smooth edges and the local order is very good. Values for average diameters were extracted by taking cross-sections of rows of dots from several such images from each sample. 
  
 Static magnetic properties of the array samples were measured using a SQuID magnetometer. Hysteresis loops were recorded at T = 295 K. Each sample displayed the double `closed loop' characteristic of vortex nucleation, movement, and annihilation in sub-micron magnetic discs. Shown in Figure \ref{s144bhys} is the loop corresponding to sample f3b, the same sample shown in the AFM image in Figure \ref{s144bAFM}. Hysteresis loops of the other samples displayed the same characteristics.
 
\begin{figure}[hbtp]
\begin{center}
\includegraphics[width=8.5cm]{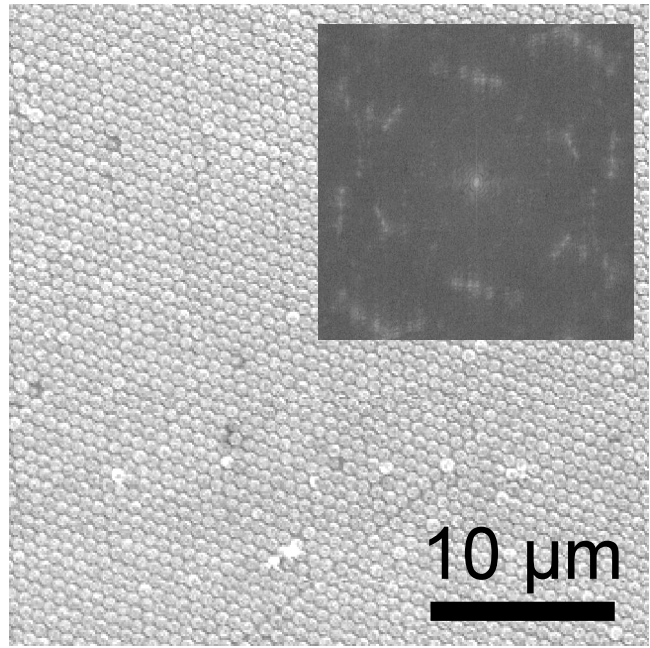}
\caption{\label{s144bFFT} An example SEM image with inset 2-D Fourier transform of a montage of 4 evenly spaced images taken over a one millimeter section of the sample: these particular images are also of sample f3b. Note that the SEM image shown was not among the images used to produce the composite FFT. Images similar to the inset from each 1 mm section of the sample were used to calculate the average variation in lattice angle per unit length, $\phi'$, for each sample.}
\end{center}

\end{figure}

Array regularity of each patterned sample was measured using Scanning Electron Microscopy (SEM). Four evenly spaced 3600$\times$ magnification images were taken from each 1 mm section of the centre line of the long axis of the film. Each of these sets of four images were stitched together to make a composite image, and the 2-dimensional Fourier transform of each composite calculated. An example of such a single SEM image with the Fourier transform of the composite image from a different section inset is shown in Figure \ref{s144bFFT}. The length of each sample was $L = 8$ mm. The Fourier transform from each 1 mm section of the sample was used to measure $\Delta \phi_i$, the variation of lattice angle over that one millimeter. The eight separate $\Delta \phi_i$ values were averaged to yield $\phi'$, the average variation in lattice angle per unit millimeter of the film: $\phi' = \sum_{i=1}^8 \Delta \phi_i / L$. $\phi'$ was used to characterize the array ordering of each patterned sample.

\begin{figure}[htbp]
\begin{center}
\includegraphics[width=8.5cm]{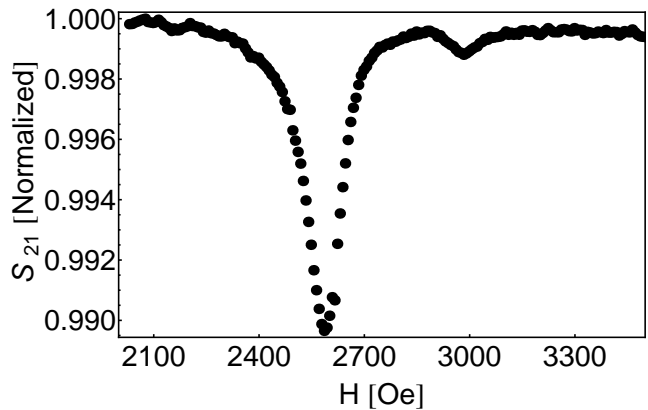}

\caption{\label{s143gs21vsH}  Plot of normalized intensity of transmission amplitude $S_{21}$ vs applied magnetic field $H$ for sample f2a at an excitation frequency of 16 GHz. In addition to the large-amplitude fundamental mode, there is a small mode at higher applied field.}
\end{center}

\end{figure}

  Each of the continuous and patterned films was characterized using microstrip-waveguide VNA-FMR. The sample was placed face-down on an 8 milli-inch copper microstrip waveguide in an in-plane saturating magnetic field. A network analyzer provided microwave excitation to the waveguide and measurement of transmission parameter $S_{21}$. Negligible reflections allowed $S_{11}$ to be ignored. \cite{Couniletal2004} In contrast to how VNA-FMR is typically performed, the excitation frequency $f$ was fixed and the applied magnetic field $H$ varied, in analogy to a cavity FMR measurement. The measurements were performed at 1 GHz intervals in the domain 7-20 GHz. A typical spectrum is shown in Figure \ref{s143gs21vsH}. Two modes can be resolved: a large amplitude mode, and a smaller amplitude mode at higher field (lower frequency). Resonance fields $H_\mathrm{res}$ and linewidths $\Delta H$ were extracted from these measurements by least-squares fitting Lorentzian curves to the troughs in $S_{21}$, using at least twenty-five data points on each side of the minimum. $f$ vs $H_\mathrm{res}$ data for continuous samples f1-3c was fitted with the Kittel equation \cite{Kittel1996} with demagnetizing factors $N_x = N_z = 0$, $N_y = 4 \pi$: 
  	 \begin{align}
 	{\omega^2}/{\gamma^2} =& (H + H_K + (N_y - N_z ) M_S ) \times \nonumber \\& (H + H_K + (N_x - N_z) M_S). \label{Kittel}
	 \end{align}
and saturation magnetizations $4 \pi M_S$ extracted. Cavity FMR measurements were performed to verify the observed mode structure.

After characterization of all samples, sample f3b was cut to obtain a smaller sample (5 $\times~5~\mathrm{mm}^2$) with a much higher degree of ordering. This sample, denoted f3b*, was then characterized in the same way as the other four samples. Additionally, FMR spectra were recorded for this sample at an excitation frequency of 10 GHz for various relative angles $\phi$ between the static applied magnetic field and the lattice vector defining the line between nearest neighbours. In addition to providing a more ordered sample, comparison of FMR results between f3b and f3b* allowed the possibility of parent film- or patterning-caused differences between array samples to be more reliably discounted.

\section{Results and Discussion}

\subsection{Film structure}
The disc diameters as measured by AFM, array ordering values as measured by SEM, and parent film magnetization values as measured by VNA-FMR are presented in Table \ref{structuretable}. The variation in disc diameter is on the order of five percent, which is the manufacturer's quoted uncertainty on the size of the reagent microspheres used and therefore the best uniformity achievable by this fabrication technique:  the disc sizes of all four samples were identical within the limits of experimental uncertainty. The narrowness of the `waist' of the SQuID hysteresis loops of the samples (Figure \ref{s144bhys}) showed that there were very few uncompensated magnetic moments at remanence, likely indicating very little roughness to the disc boundaries.

Each of the five patterned samples was distinguished by a unique ordering parameter $\phi'$, although there was some overlap between f3a and f2a at the limits of experimental uncertainty. Each of the three parent films' saturation magnetizations $M_S$ were slightly different. The linewidths of these films at a given frequency, $\Delta H$ were also different. For this reason, VNA-FMR data for the parent continuous films f3, f3, and f1 are included in the tables in this section. In Figure \ref{fwhmvsf} the linewidths for the parent continuous films are presented alongside those of the corresponding patterned films. In the comparison of inhomogeneous broadenings between films shown in Figure \ref{delHvsPhi} these parent film linewidths have been accounted for.

\begin{table}[h]
\begin{center}
\begin{tabular}{ | c || c   c   c  | }
    \hline
    Film & d [nm] & $\phi' [{}^{\circ}~\mathrm{mm^{-1}}$]   & $4 \pi M_S$ [kOe]  \\ \hline \hline
    f3c & - & - & 8.49 \\
    f3b* & $695 \pm 28$ & $2.6 \pm 0.4$ & -\\
    f3b & $695 \pm 28$ & $6.0 \pm 0.8$ & - \\ 
   f3a &$ 703 \pm 37$ &  $9.4 \pm 1.1$ & -  \\
   f2c & - & - & 8.69 \\
  f2a & $697 \pm 31$ & $11.3 \pm 1.7$ & -  \\
  f1c & - & - & 8.85 \\
   f1a & $ 699 \pm 28$ & $19.9 \pm 2.1$ & -  \\
    \hline
\end{tabular}
\end{center}
\caption{Table showing average disc diameter, $d$, array variation per unit length $\phi'$, and saturation magnetization $4 \pi M_S$ for the samples used in this study.}
\label{structuretable}
\end{table}

\subsection{FMR mode structure}

A similar mode to the high-field mode in Figure \ref{s143gs21vsH} has been calculated to exist for isolated discs of larger aspect ratio $L/d$ that those fabricated for this study, using OOMMF\cite{Shawetal2009} and linearized micromagnetic approaches.\cite{RivkinetalPRB2007, Giovanninietal2007} In these publications the mode is calculated to be located in the `end' of the disc as defined by the direction external magnetic field. These two studies also report experimental measurements of these modes in closely packed disc systems. Such modes have also been measured elsewhere in systems of comparable disc dimensions to those fabicated for this study. \cite{Nevirkovetsetal2008, Jungetalexp2002} VNA-FMR lacks the kind of spatial resolution of techniques like magnetic resonance force microscopy, and in this study the possible confinement of the high-field mode to specific regions of the discs could not be investigated directly.

\begin{figure}[htbp]
\begin{center}
\includegraphics[width=8.5cm]{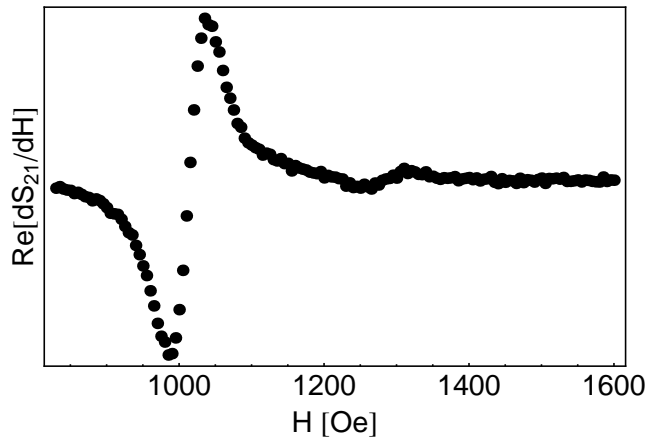}

\caption{\label{s141icavfmr}  Plot of the real part of ${d S_{21}}/{d H}$ vs $H$ for sample f1a at an excitation frequency of 9.55 GHz, as measured by cavity FMR. The high-field mode seen in Figure \ref{s143gs21vsH} is preserved in the uniform excitation field of the FMR cavity. No units are included on the vertical axes because the scale is dependent on the particular tuning of the cavity.}
\end{center}

\end{figure}

Several possibilities for the origin of the mode were ruled out experimentally. The possibility that the mode was caused by a non-uniform excitation field resulting from microwave screening by eddy currents in the sample\cite{Kostylev2009} was ruled out by measurement of the mode structure by cavity FMR. In an FMR cavity, the excitation field is very uniform. Figure \ref{s141icavfmr} shows the cavity FMR spectrum. The higher field mode is still present, indicating that it does not result from non-uniform microwave excitation. The amplitude of this mode as measured by VNA-FMR did not vary significantly between samples of different degrees of ordering. This ruled out the possibility of the high-field mode being the result of a collective excitation occurring at discontinuities or around vacancies in the array, which are more numerous in more disordered samples. Finally, the out-of-plane saturated FMR spectrum--not included here for brevity--showed up to five well resolved modes in the structure expected for cylindrical symmetry, \cite{Kakazeietal2004} suggesting that the high-field mode was not the product of some repeated non-cylindrical element. Remaining possible explanations for the high-field mode are that it is an `end' or `edge' mode as in References \onlinecite{Jungetalexp2002},  \onlinecite{RivkinetalPRB2007}, and \onlinecite{Shawetal2009}, or that it arises out of or is modified by some collective effect of the array itself.

The relationship between the resonance field $H_\mathrm{res}$ of both modes and the relative lattice-applied field angle $\phi$ was measured for sample f3b*. Neither mode showed any angular variation beyond the the 15 and 20 Oe scatter for the fundamental and high-field modes, respectively. This scatter was caused by a combination of small movements of the sample in the applied magnetic field when varying $\phi$ and uncertainty in the baseline of the Lorentzian fit used to extract the values of $H_\mathrm{res}$. This measurement restricts the angular variation of $H_\mathrm{res}$ to be below these values of uncertainty. Additionally, the amplitudes of both modes were essentially independent of $\phi$.

In previous studies by other authors on square arrays of sub-micron discs of similar diameter-to-pitch ratio, strong dependencies on the relative array-field angle $\phi$ of the resonance field have $H_\mathrm{res}$ have been measured for both fundamental and high-field modes. \cite{Jungetalexp2002, Kakazeietal2006, RivkinetalPRB2007} Given the higher symmetry of the trigonal system, and the large diameter of the discs with respect to the pitch, it is not entirely surprising that such strong dependence was not observed in the fundamental mode of the arrays fabricated for this study.

 \begin{figure}[htbp]
\begin{center}

\includegraphics[width=8.5cm]{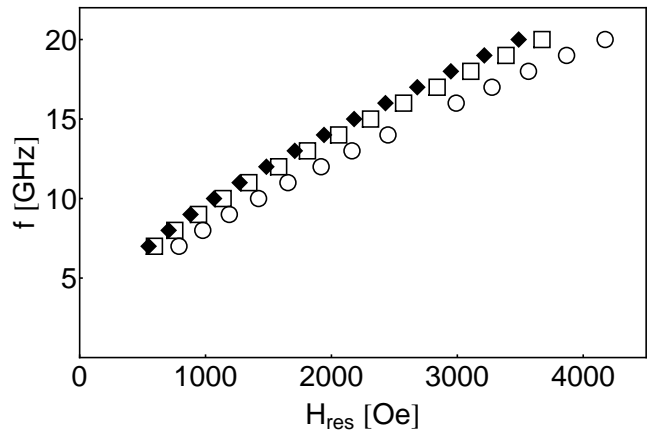}

\caption{\label{s141fvsH}  Plot of resonance frequency $f$ against resonance field $H_{\mathrm{res}}$ for sample f1a (squares; circles - high-field mode) and f1c (diamonds), a sample of the continuous film from which f1a was patterned, as measured by VNA-FMR. No resonances were observed below the saturating field of the dots. }
\end{center}

\end{figure}

 The $f$ vs $H_\mathrm{res}$ data for film f1a is shown in Figure \ref{s141fvsH}; this data was representative of equivalent measurements of the other four samples. Resonances below 7 GHz were not recorded, so as to ensure that the resonance field was always high enough for the discs to be tangentially magnetically saturated. The data for each of the two modes for each sample was fit with Equation \ref{Kittel}, under the assumption $N_z = N_x$, using $N_y-N_z$ as a fit parameter. The results of this process are tabulated in Table \ref{Nmtableip}. There was no clear variation with $\phi'$ in $N_y-N_z$ for either mode across the samples.
 
\begin{table}[h]
\begin{center}
\begin{tabular}{ | c || c   c    | }
    \hline
    Film & $N_y - N_z$ fundamental & $N_y-N_z$ high-field   \\ \hline \hline
    f3b* & $0.958 \pm 0.001$ & $0.740 \pm 0.002$ \\
    f3b & $0.934 \pm 0.001$ & $0.716 \pm 0.001$  \\ 
   f3a &$ 0.939 \pm 0.001$ &  $0.720 \pm 0.002$   \\
  f2a & $0.947 \pm 0.001$ & $0.724 \pm 0.001$  \\
   f1a & $ 0.937 \pm 0.001$ & $0.709 \pm 0.002$   \\
    \hline
\end{tabular}
\end{center}
\caption{Table showing fit parameters $N_y$ for the fundamental and high-field modes, as extracted by fitting Equation \ref{Kittel}. }
\label{Nmtableip}
\end{table}

Significant differences between the resonance frequencies of close-packed and well-isolated discs have been reported for square arrays in one other paper. \cite{Jorzicketal1999} This effect was not observed in this study for the case where array ordering--rather than array pitch--was varied: there appears to be no correlation for this sample series between array order $\phi'$ and fitted demagnetizing parameter $N_y-N_z$.

\subsection{FMR linewidth}
 
\begin{figure*}[htbp]
\begin{center}

\includegraphics[width=17cm]{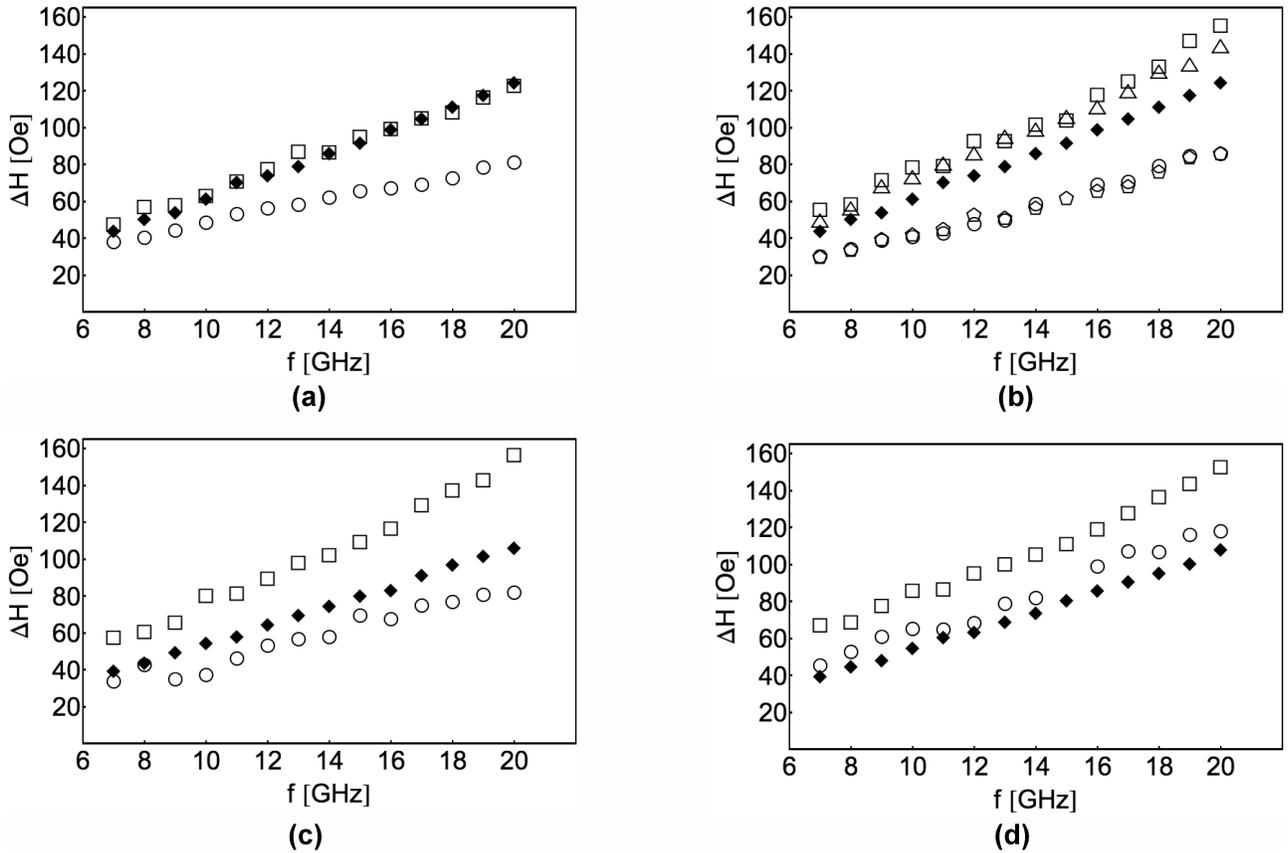}

\caption{\label{fwhmvsf}  Plot of full-width-half maximum linewidth $\Delta H$ against resonance frequency $f$ as measured by VNA-FMR for samples (a) f3b*; (b) f3b (open triangles: fundamental mode; open pentagons: high-field mode); (c) f2a; and (d) f1a. The filled diamonds are the data points for the parent continuous films. For all samples except f3b, the open squares denote the fundamental mode, and the open circles the high-field mode of the patterned films listed; the key for f3b has already been listed in parentheses. }
\end{center}

\end{figure*}

The full-width half-maximum linewidth values $\Delta H$ extracted from Lorentzian fits to the FMR modes are shown in Figure  \ref{fwhmvsf}. Each plot shows the $\Delta H$ vs $f$ data for the unpatterned parent film, and both the fundamental and high-field modes for the patterned films. The uncertainties in these linewidth values are not shown on the graphs for clarity: for the fundamental mode these uncertainties were on the order of $\pm 2\%$, and for the high-field mode $\pm 20\%$. In both cases the primary source was uncertainty in the baseline of the Lorentzian fit; the uncertainty was more severe for the high-field mode because its smaller amplitude and proximity to the fundamental mode made the value of the linewidth highly dependent on the number of points on either side of the peak used in the fit. It is clear from examination that the linewidths of both the fundamental and high-field modes increase with increasing array disorder.

\begin{table}[h]
\begin{center}
\begin{tabular}{ | c  c  || c   c   c  | }
    \hline
    Film & $\phi' [{}^{\circ}~\mathrm{mm^{-1}}$] & $\alpha \times 10^3$  & $\Delta H_0$ [Oe] & $\Delta H_{0,~\alpha}$ [Oe] \\ \hline \hline
    f3c & - & $9.28 \pm 0.16$ & $0.25 \pm 1.20$ & - \\
    f3b* & $2.6 \pm 0.4$ & $8.61 \pm 0.16$ & $8.24 \pm 1.26$ & $2.31 \pm 3.10$ \\
    f3b & $6.0 \pm 0.8$ & $10.70 \pm 0.18$ & $1.02 \pm 1.38$ & $13.68 \pm 3.33$ \\ 
   f3a & $9.4 \pm 1.1$ & $10.84 \pm 0.19$  & $3.54 \pm 1.48$ & $17.89 \pm 3.41$ \\
    f2c & - & $7.79 \pm 0.14$ & $2.48 \pm 1.06$ & - \\
   f2a & $11.3 \pm 1.7$ & $11.00 \pm 0.19$  & $3.27 \pm 1.50$ & $32.21 \pm 3.25$ \\
     f1c & - & $7.77 \pm 0.14$ & $2.85 \pm 1.06$ & - \\
   f1a & $19.9 \pm 2.1$ & $9.55 \pm 0.21$& $19.90 \pm 1.64$ & $36.08 \pm 3.33$ \\
    \hline
\end{tabular}
\end{center}
\caption{Table showing fit data for each sample to Equation \ref{Silvafwhmeqn} for the fundamental mode data, for the case where $\alpha$ is allowed to vary for patterned films and for the case when it is held to the value $\alpha_\mathrm{cnts}$ of the corresponding continuous film.}
\label{orderingdata}
\end{table}

\begin{table}[h]
\begin{center}
\begin{tabular}{ | c  c  || c   c  | }
    \hline
    Film & $\phi' [{}^{\circ}~\mathrm{mm^{-1}}$] & $\alpha \times 10^3$  & $\Delta H_0$ [Oe]  \\ \hline \hline
    f3b* & $2.6 \pm 0.4$ & $5.00 \pm 1.14$ & $14.74 \pm 9.23$ \\ 
    f3b & $6.0 \pm 0.8$ & $6.21 \pm 1.06$ & $0.95 \pm 8.21$ \\ 
   f3a & $9.4 \pm 1.1$ & $6.40 \pm 1.11$  & $-0.92 \pm 8.36$  \\
  f2a & $11.3 \pm 1.7$ & $6.10 \pm 1.12$  & $2.80 \pm 8.70$  \\
   f1a & $19.9 \pm 2.1$ & $8.45 \pm 1.61$& $6.68 \pm 12.37$  \\
    \hline
\end{tabular}
\end{center}
\caption{Table showing fit data for each sample to Equation \ref{Silvafwhmeqn} for the high-field mode data. The large uncertainties of the fit parameters are the direct result of an estimated 20\% error on high-field mode linewidth data.}
\label{orderingdataedge}
\end{table}

In the context of studies of spin-wave mode broadening, Shaw \emph{et al.} have pointed out the importance of separating the effects of intrinsic damping from inhomogeneous damping \cite{Kalarickaletal2006, Shawetal2009} in the total linewidth $\Delta H$, as defined by the equation:
 \begin{align}
\Delta H &= \Delta H_0 + {4 \pi \alpha}/({\gamma \mu_0} f) ~~~~\mathrm{(SI).}
\label{Silvafwhmeqn}
\end{align}
 Here $f$ is the frequency of precession and $\gamma$ the gyromagnetic ratio. $\alpha$ is the intrinsic damping parameter in the Landau-Lifshitz-Gilbert equation, \cite{LandauandLifshitz1935, GilbertPhDThesis} and can be thought of as a `viscous' damping of energy to the lattice. \cite{Lenzetal2006}
 
 $\Delta H_0$ is a term representing inhomogeneous broadening: an example of such a broadening might be the decay of uniform motion into spin waves having non-zero wavevectors, which might themselves decay to the lattice.\cite{Bloch1946} The $\Delta H$ vs $f$ data represented in Figure \ref{fwhmvsf} (a)-(d) were fit with equation \ref{Silvafwhmeqn}. The extracted fit parameters $\alpha$ and $\Delta H_0$ are given in Table \ref{orderingdata} for the fundamental modes and Table \ref{orderingdataedge} for the high-field modes.
  
  Previous studies by other authors of FMR linewidth broadening in disc-geometry nanostructures \cite{Sankeyetal2006, Schneideretal2007, Shawetal2009} have produced some conflicting results as to whether patterning affects the value of the intrinsic damping parameter, $\alpha$. In the study presented here, $\alpha$ values did appear to change outside the bounds of experimental uncertainty. For the fundamental mode as measured in this study, $\alpha$ was larger for the patterned samples than for their corresponding parent continuous films; the exception was the most ordered sample, for which $\alpha$ was slightly lower. Conversely, $\alpha$ for the high-field mode was lower than or within the range of experimental error of the corresponding value for the fundamental mode. Although the values of $\alpha$ for patterned films appeared to be different than for the corresponding parent films, there was no clear variation of $\alpha$ for either mode with the parameter $\phi'$ which distinguished the films.

 It is difficult to interpret these $\alpha$ values because of the lack of a complete theoretical description in the literature of spin wave modes in closely packed and potentially dipole coupled tangentially magnetized disc arrays. This lack is a consequence of the difficulties caused by the absence of cylindrical symmetry in the plane of the discs, \cite{Gubbiottietal2003, GuslienkoandSlavin2000} and complicated for this experimental system by the lack of perfect ordering. For these reasons, it is difficult to know that Equation \ref{Silvafwhmeqn} holds in this system. In particular, one can speculate that the inhomogenous broadening $\Delta H_0$ may have a frequency- or field-dependence which varies with array ordering, in analogy to or even contributed to by the field dependence of extrinsic damping related to two-magnon scattering processes. \cite{AriasandMills1999} Such dependency might go some way to explaining the apparent differences in $\alpha$ between patterned and unpatterned films. If $\Delta H_0$ is a function of field (or frequency), then the steeper gradients in $\Delta H$ seen in Figure \ref{fwhmvsf} might not require the `real' intrinsic damping parameter to differ between patterned and parent films.

\begin{figure}[htbp]
\begin{center}

\includegraphics[width=8.5cm]{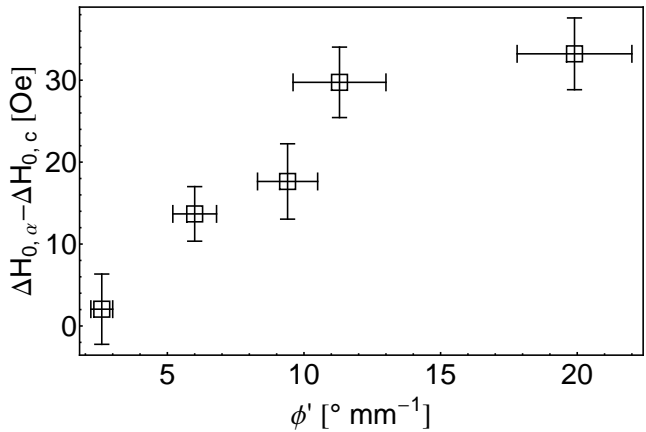}

\caption{\label{delHvsPhi}  Plot of $\Delta H_{0,~\alpha} - \Delta H_{0, \mathrm{c}}$ vs $\phi'$ for the fundamental mode data of patterned films f3b*, f3b, f3a, f2a, and f1a. The data plotted here corresponds to the appropriate data presented in tabular format in Table \ref{orderingdata}. }
\end{center}

\end{figure}

  Since the origin of these differing gradients was not clear, a second set of fits to Equation \ref{Silvafwhmeqn} was performed,  with the slopes of the patterned film data set to be equal to those of the parent continuous films: $\alpha = \alpha_\mathrm{c}$. The difference $\Delta H_{0,~\alpha} - \Delta H_{0,~\mathrm{c}}$ between this `fixed intrinsic damping inhomogeneous broadening' value $\Delta H_{0,~\alpha}$ and the continuous film inhomogeneous broadening, $\Delta H_{0,~\mathrm{c}}$, is equivalent to the average of the difference between linewidths $\Delta H$ of the patterned and unpatterned films over the range of frequencies measured. It is a convenient experimental index of the inhomogeneous broadening that takes into account possible changes to $\alpha$ in the frequency domain of interest, rather than a quantitative measure of the `true' inhomogeneous broadening as defined by Equation \ref{Silvafwhmeqn}. A plot of this difference of inhomogeneous broadening, $\Delta H_{0,~\alpha} - \Delta H_{0,~\mathrm{c}}$, vs ordering parameter $\phi'$ is shown in Figure \ref{delHvsPhi}. The increase of inhomogeneous broadening with increasing array disorder can be clearly seen.
 
  The difficulties encountered in fitting the $\Delta H_0$ vs $f$ data were even more pronounced for the high-field mode. Since this mode is unique to the patterned geometry, there was no analogue from the parent continuous films with which to compare data. Additionally, there was a high degree of experimental uncertainty in high-field mode $\Delta H$ values, which led to high uncertainties in $\alpha$. This in turn produced $\Delta H_0$ uncertainties that were comparable to or greater than the $\Delta H_0$ values themselves. Despite these uncertainties, the data in Figure \ref{fwhmvsf} suggest that the linewidth for this mode also increases with increasing $\phi'$ in the frequency range measured.
  
 Disc uniformity has been maintained as far as experimentally practicable, as shown by consistency of hysteretic measurements and AFM images. Differences in linewidth and saturation magnetization of parent films have been accounted for in the comparison between patterned films. The variation in broadening with changing array ordering suggests that the discs are dipole coupled together, and that the degree of coupling affects the FMR linewidth of both modes. That sample f3b* shows a more narrow linewidth than the patterned film f3b from which it was cut supports this conjecture. It might have been speculated that differently arranged array domains in the more disordered samples, interacting strongly internally but weakly with other domains, were producing different resonance fields due to angular dependence of $H_\mathrm{res}$. However, the lack of angular dependence of $H_\mathrm{res}$ in either mode in sample f3b* suggests that the broadening is not due to the sum of differing resonance fields of different array domains, but rather to some less local effect.

In the only previous study of the FMR characteristics of trigonal arrays fabricated by a self-assembly process, it was not clear whether the anomalously large linewidth recorded was due to the fabrication process, a distribution of disc parameters, or the quality of the parent continuous film. \cite{Kostylevetal2008IEEE} The results of the present study indicate that there is no dramatic impact on linewidth due to patterning in such a process: even for the most disordered array, f1a, there was less than a factor-of-two broadening;  for the case of near-perfect order represented by sample f3b*, there was essentially no broadening. The broad linewidth seen in Reference \onlinecite{Kostylevetal2008IEEE} is therefore likely to have been caused by a broad linewidth in the parent film rather than the patterning process itself. The narrowness of the `waist' of the vortex hysteresis loops and high magnetization of the patterned films studied here are consistent with this conclusion. The utility of the water-surface self-assembly lithographic process for patterning disc arrays has thus been partially vindicated: despite the reasonable criticisms that the self-assembly process does not preserve long range order, the impact of variations in that order on the FMR linewidth have been shown to be relatively minor.

\section{Conclusions}

A series of arrays of sub-micron Permalloy discs was fabricated by a water-surface self-assembly lithographic process. The geometry, static magnetic behavior, and ordering of the disc arrays were measured. The arrays were characterized by vector network analyzer ferromagnetic resonance. Two modes were present in the ferromagnetic resonance spectra: a fundamental mode and a higher field, lower frequency mode. The variation of the resonance field with relative lattice-to-static field angle was shown to be negligible for both modes. The relationship between resonance frequency and resonance field was found to be uncorrelated to increasing array disorder. The linewidth vs frequency data for these arrays revealed that the self-assembly fabrication process did not dramatically increase the ferromagnetic resonance linewidth, clarifying questions from a previous study. The intrinsic damping parameter for the fundamental mode appeared to be higher for the patterned films than in the parent continuous films, but it was not clear whether this was a result of field- or frequency-dependencies of the inhomogeneous damping not accounted for in the fit. The linewidth of both the fundamental and high-field modes was found to increase with increasing array disorder.

\section*{Acknowledgements}

This work was supported in part by the Australian Research Council under Discovery Grant ``Magnetic nanostructures for emerging technologies". N. Ross is supported by a University of Western Australia Hackett Postgraduate Scholarship. The authors acknowledge the facilities, scientific and technical assistance of the Australian Microscopy \& Microanalysis Research Facility at the Centre for Microscopy, Characterisation \& Analysis, The University of Western Australia, a facility funded by The University, State and Commonwealth Governments. The authors are grateful to the UWA Microelectronics Research Group, Biomagnetics Group, and Murray Baker Research Group for generous provision of experimental facilities; to  M. Madami for helpful discussions regarding OOMMF software; and to F. Y. Ogrin and E. Sirotkin for their assistance and helpful discussions regarding water-surface nanosphere lithography.

%


\begin{thebibliography}{10}%
\makeatletter
\providecommand \@ifxundefined [1]{%
 \ifx #1\undefined \expandafter \@firstoftwo
 \else \expandafter \@secondoftwo
\fi
}%
\providecommand \@ifnum [1]{%
 \ifnum #1\expandafter \@firstoftwo
 \else \expandafter \@secondoftwo
\fi
}%
\providecommand \enquote [1]{``#1''}%
\providecommand \bibnamefont  [1]{#1}%
\providecommand \bibfnamefont [1]{#1}%
\providecommand \citenamefont [1]{#1}%
\providecommand\href[0]{\@sanitize\@href}%
\providecommand\@href[1]{\endgroup\@@startlink{#1}\endgroup\@@href}%
\providecommand\@@href[1]{#1\@@endlink}%
\providecommand \@sanitize [0]{\begingroup\catcode`\&12\catcode`\#12\relax}%
\@ifxundefined \pdfoutput {\@firstoftwo}{%
 \@ifnum{\z@=\pdfoutput}{\@firstoftwo}{\@secondoftwo}%
}{%
 \providecommand\@@startlink[1]{\leavevmode}%
 \providecommand\@@endlink[0]{}%
}{%
 \providecommand\@@startlink[1]{%
  \leavevmode
  \pdfstartlink
   attr{/Border[0 0 1 ]/H/I/C[0 1 1]}%
   user{/Subtype/Link/A<</Type/Action/S/URI/URI(#1)>>}%
  \relax
 }%
 \providecommand\@@endlink[0]{\pdfendlink}%
}%
\providecommand \url  [0]{\begingroup\@sanitize \@url }%
\providecommand \@url [1]{\endgroup\@href {#1}{\urlprefix}}%
\providecommand \urlprefix [0]{URL }%
\providecommand \Eprint[0]{\href }%
\@ifxundefined \urlstyle {%
  \providecommand \doi [1]{doi:\discretionary{}{}{}#1}%
}{%
  \providecommand \doi [0]{doi:\discretionary{}{}{}\begingroup
  \urlstyle{rm}\Url }%
}%
\providecommand \doibase [0]{http://dx.doi.org/}%
\providecommand \Doi[1]{\href{\doibase#1}}%
\providecommand \selectlanguage [0]{\@gobble}%
\providecommand \bibinfo [0]{\@secondoftwo}%
\providecommand \bibfield [0]{\@secondoftwo}%
\providecommand \translation [1]{[#1]}%
\providecommand \BibitemOpen[0]{}%
\providecommand \bibitemStop [0]{}%
\providecommand \bibitemNoStop [0]{.\EOS\space}%
\providecommand \EOS [0]{\spacefactor3000\relax}%
\providecommand \BibitemShut [1]{\csname bibitem#1\endcsname}%
\bibitem{TerrisandThomson2005}%
  \BibitemOpen
  \bibfield{author}{%
  \bibinfo {author} {\bibfnamefont{B.~D.}\ \bibnamefont{Terris}}\ and\ \bibinfo
  {author} {\bibfnamefont{T.}~\bibnamefont{Thomson}},\ }%
  \bibfield{journal}{%
  \bibinfo {journal} {Journal of Physics D}\ }%
  \textbf{\bibinfo {volume} {38}},\ \bibinfo {pages} {R199} (\bibinfo {year}
  {2005})\BibitemShut{NoStop}%
\bibitem{Thomsonetal2006}%
  \BibitemOpen
  \bibfield{author}{%
  \bibinfo {author} {\bibfnamefont{T.}~\bibnamefont{Thomson}}, \bibinfo
  {author} {\bibfnamefont{G.}~\bibnamefont{Hu}},\ and\ \bibinfo {author}
  {\bibfnamefont{B.~D.}\ \bibnamefont{Terris}},\ }%
  \bibfield{journal}{%
  \bibinfo {journal} {Physical Review Letters}\ }%
  \textbf{\bibinfo {volume} {96}},\ \bibinfo {pages} {257204} (\bibinfo {year}
  {2006})\BibitemShut{NoStop}%
\bibitem{Shawetal2007}%
  \BibitemOpen
  \bibfield{author}{%
  \bibinfo {author} {\bibfnamefont{J.~M.}\ \bibnamefont{Shaw}}, \bibinfo
  {author} {\bibfnamefont{W.~H.}\ \bibnamefont{Rippard}}, \bibinfo {author}
  {\bibfnamefont{S.~E.}\ \bibnamefont{Russek}}, \bibinfo {author}
  {\bibfnamefont{T.}~\bibnamefont{Reith}},\ and\ \bibinfo {author}
  {\bibfnamefont{C.~M.}\ \bibnamefont{Falco}},\ }%
  \bibfield{journal}{%
  \bibinfo {journal} {Journal of Applied Physics}\ }%
  \textbf{\bibinfo {volume} {101}},\ \bibinfo {pages} {023909} (\bibinfo {year}
  {2007})\BibitemShut{NoStop}%
\bibitem{Hellwigetal2007}%
  \BibitemOpen
  \bibfield{author}{%
  \bibinfo {author} {\bibfnamefont{O.}~\bibnamefont{Hellwig}}, \bibinfo
  {author} {\bibfnamefont{A.}~\bibnamefont{Berger}}, \bibinfo {author}
  {\bibfnamefont{T.}~\bibnamefont{Thomson}}, \bibinfo {author}
  {\bibfnamefont{E.}~\bibnamefont{Dobisz}}, \bibinfo {author}
  {\bibfnamefont{Z.~Z.}\ \bibnamefont{Bandic}}, \bibinfo {author}
  {\bibfnamefont{H.}~\bibnamefont{Yang}}, \bibinfo {author}
  {\bibfnamefont{D.~S.}\ \bibnamefont{Kercher}},\ and\ \bibinfo {author}
  {\bibfnamefont{E.~E.}\ \bibnamefont{Fullerton}},\ }%
  \bibfield{journal}{%
  \bibinfo {journal} {Applied Physics Letters}\ }%
  \textbf{\bibinfo {volume} {90}},\ \bibinfo {pages} {162516} (\bibinfo {year}
  {2007})\BibitemShut{NoStop}%
\bibitem{Shawetal2008}%
  \BibitemOpen
  \bibfield{author}{%
  \bibinfo {author} {\bibfnamefont{J.~M.}\ \bibnamefont{Shaw}}, \bibinfo
  {author} {\bibfnamefont{S.~E.}\ \bibnamefont{Russek}}, \bibinfo {author}
  {\bibfnamefont{T.}~\bibnamefont{Thomson}}, \bibinfo {author}
  {\bibfnamefont{M.~J.}\ \bibnamefont{Donahue}}, \bibinfo {author}
  {\bibfnamefont{B.~D.}\ \bibnamefont{Terris}}, \bibinfo {author}
  {\bibfnamefont{O.}~\bibnamefont{Hellwig}}, \bibinfo {author}
  {\bibfnamefont{E.}~\bibnamefont{Dobisz}},\ and\ \bibinfo {author}
  {\bibfnamefont{M.~L.}\ \bibnamefont{Schneider}},\ }%
  \bibfield{journal}{%
  \bibinfo {journal} {Physical Review B}\ }%
  \textbf{\bibinfo {volume} {78}},\ \bibinfo {pages} {024414} (\bibinfo {year}
  {2008})\BibitemShut{NoStop}%
\bibitem{Kiselevetal2003}%
  \BibitemOpen
  \bibfield{author}{%
  \bibinfo {author} {\bibfnamefont{S.~I.}\ \bibnamefont{Kiselev}}, \bibinfo
  {author} {\bibfnamefont{J.~C.}\ \bibnamefont{Sankey}}, \bibinfo {author}
  {\bibfnamefont{I.~N.}\ \bibnamefont{Krivorotov}}, \bibinfo {author}
  {\bibfnamefont{N.~C.}\ \bibnamefont{Emley}}, \bibinfo {author}
  {\bibfnamefont{R.~J.}\ \bibnamefont{Schoelkopf}}, \bibinfo {author}
  {\bibfnamefont{R.~A.}\ \bibnamefont{Buhrman}},\ and\ \bibinfo {author}
  {\bibfnamefont{D.~C.}\ \bibnamefont{Ralph}},\ }%
  \bibfield{journal}{%
  \bibinfo {journal} {Nature}\ }%
  \textbf{\bibinfo {volume} {425}},\ \bibinfo {pages} {380} (\bibinfo {year}
  {2003})\BibitemShut{NoStop}%
\bibitem{Deacetal2008}%
  \BibitemOpen
  \bibfield{author}{%
  \bibinfo {author} {\bibfnamefont{A.~M.}\ \bibnamefont{Deac}}, \bibinfo
  {author} {\bibfnamefont{A.}~\bibnamefont{Fukushima}}, \bibinfo {author}
  {\bibfnamefont{H.}~\bibnamefont{Kubota}}, \bibinfo {author}
  {\bibfnamefont{H.}~\bibnamefont{Maehara}}, \bibinfo {author}
  {\bibfnamefont{Y.}~\bibnamefont{Suzuki}}, \bibinfo {author}
  {\bibfnamefont{S.}~\bibnamefont{Yuasa}}, \bibinfo {author}
  {\bibfnamefont{Y.}~\bibnamefont{Nagamine}}, \bibinfo {author}
  {\bibfnamefont{K.}~\bibnamefont{Tsunekawa}}, \bibinfo {author}
  {\bibfnamefont{D.~D.}\ \bibnamefont{Djayaprawira}},\ and\ \bibinfo {author}
  {\bibfnamefont{N.}~\bibnamefont{Watanabe}},\ }%
  \bibfield{journal}{%
  \bibinfo {journal} {Nature Physics}\ }%
  \textbf{\bibinfo {volume} {4}},\ \bibinfo {pages} {803} (\bibinfo {year}
  {2008})\BibitemShut{NoStop}%
\bibitem{Zabowetal2008}%
  \BibitemOpen
  \bibfield{author}{%
  \bibinfo {author} {\bibfnamefont{G.}~\bibnamefont{Zabow}}, \bibinfo {author}
  {\bibfnamefont{S.}~\bibnamefont{Dodd}}, \bibinfo {author}
  {\bibfnamefont{J.}~\bibnamefont{Moreland}},\ and\ \bibinfo {author}
  {\bibfnamefont{A.}~\bibnamefont{Koretsky}},\ }%
  \bibfield{journal}{%
  \bibinfo {journal} {Nature}\ }%
  \textbf{\bibinfo {volume} {453}},\ \bibinfo {pages} {1058} (\bibinfo {year}
  {2008})\BibitemShut{NoStop}%
\bibitem{Kimetal2009}%
  \BibitemOpen
  \bibfield{author}{%
  \bibinfo {author} {\bibfnamefont{D.}~\bibnamefont{Kim}}, \bibinfo {author}
  {\bibfnamefont{E.~A.}\ \bibnamefont{Rozhkova}}, \bibinfo {author}
  {\bibfnamefont{I.~V.}\ \bibnamefont{Ulasov}}, \bibinfo {author}
  {\bibfnamefont{S.~D.}\ \bibnamefont{Bader}}, \bibinfo {author}
  {\bibfnamefont{R.}~\bibnamefont{Rajh}}, \bibinfo {author}
  {\bibfnamefont{M.~S.}\ \bibnamefont{Lesniak}},\ and\ \bibinfo {author}
  {\bibfnamefont{V.}~\bibnamefont{Novosad}},\ }%
  \bibfield{journal}{%
  \bibinfo {journal} {Nature Materials}\ }%
  \textbf{\bibinfo {volume} {9}},\ \bibinfo {pages} {165} (\bibinfo {year}
  {2010})\BibitemShut{NoStop}%
\bibitem{Kakazeietal2006}%
  \BibitemOpen
  \bibfield{author}{%
  \bibinfo {author} {\bibfnamefont{G.~N.}\ \bibnamefont{Kakazei}}, \bibinfo
  {author} {\bibfnamefont{Y.~G.}\ \bibnamefont{Pogorelov}}, \bibinfo {author}
  {\bibfnamefont{M.~D.}\ \bibnamefont{Costa}}, \bibinfo {author}
  {\bibfnamefont{T.}~\bibnamefont{Mewes}}, \bibinfo {author}
  {\bibfnamefont{P.~E.}\ \bibnamefont{Wigen}}, \bibinfo {author}
  {\bibfnamefont{P.~C.}\ \bibnamefont{Hammel}}, \bibinfo {author}
  {\bibfnamefont{V.~O.}\ \bibnamefont{Golub}}, \bibinfo {author}
  {\bibfnamefont{T.}~\bibnamefont{Okuno}},\ and\ \bibinfo {author}
  {\bibfnamefont{V.}~\bibnamefont{Novosad}},\ }%
  \bibfield{journal}{%
  \bibinfo {journal} {Physical Review B}\ }%
  \textbf{\bibinfo {volume} {74}},\ \bibinfo {pages} {060406(R)} (\bibinfo
  {year} {2006})\BibitemShut{NoStop}%
\bibitem{Kostylevetal2008IEEE}%
  \BibitemOpen
  \bibfield{author}{%
  \bibinfo {author} {\bibfnamefont{M.}~\bibnamefont{Kostylev}}, \bibinfo
  {author} {\bibfnamefont{R.}~\bibnamefont{Magaraggia}}, \bibinfo {author}
  {\bibfnamefont{F.~Y.}\ \bibnamefont{Ogrin}}, \bibinfo {author}
  {\bibfnamefont{E.}~\bibnamefont{Sirotkin}}, \bibinfo {author}
  {\bibfnamefont{V.~F.}\ \bibnamefont{Mescheryakov}}, \bibinfo {author}
  {\bibfnamefont{N.}~\bibnamefont{Ross}},\ and\ \bibinfo {author}
  {\bibfnamefont{R.~L.}\ \bibnamefont{Stamps}},\ }%
  \bibfield{journal}{%
  \bibinfo {journal} {IEEE Transactions on Magnetics}\ }%
  \textbf{\bibinfo {volume} {44}},\ \bibinfo {pages} {2741} (\bibinfo {year}
  {2008})\BibitemShut{NoStop}%
\bibitem{Nevirkovetsetal2008}%
  \BibitemOpen
  \bibfield{author}{%
  \bibinfo {author} {\bibfnamefont{I.~P.}\ \bibnamefont{Nevirkovets}}, \bibinfo
  {author} {\bibfnamefont{O.}~\bibnamefont{Chernyashevskyy}}, \bibinfo {author}
  {\bibfnamefont{J.~B.}\ \bibnamefont{Ketterson}}, \bibinfo {author}
  {\bibfnamefont{V.}~\bibnamefont{Metlushko}},\ and\ \bibinfo {author}
  {\bibfnamefont{B.~K.}\ \bibnamefont{Sarma}},\ }%
  \bibfield{journal}{%
  \bibinfo {journal} {Journal of Applied Physics}\ }%
  \textbf{\bibinfo {volume} {104}},\ \bibinfo {pages} {063920} (\bibinfo {year}
  {2008})\BibitemShut{NoStop}%
\bibitem{Tsaietal2009}%
  \BibitemOpen
  \bibfield{author}{%
  \bibinfo {author} {\bibfnamefont{C.~C.}\ \bibnamefont{Tsai}}, \bibinfo
  {author} {\bibfnamefont{J.}~\bibnamefont{Choi}}, \bibinfo {author}
  {\bibfnamefont{S.}~\bibnamefont{Cho}}, \bibinfo {author}
  {\bibfnamefont{S.~J.}\ \bibnamefont{Lee}}, \bibinfo {author}
  {\bibfnamefont{B.~K.}\ \bibnamefont{Sarma}}, \bibinfo {author}
  {\bibfnamefont{C.}~\bibnamefont{Thompson}}, \bibinfo {author}
  {\bibfnamefont{O.}~\bibnamefont{Chernyashvskyy}}, \bibinfo {author}
  {\bibfnamefont{I.}~\bibnamefont{Nevirkovets}}, \bibinfo {author}
  {\bibfnamefont{V.}~\bibnamefont{Metlushko}}, \bibinfo {author}
  {\bibfnamefont{K.}~\bibnamefont{Rivkin}},\ and\ \bibinfo {author}
  {\bibfnamefont{J.~B.}\ \bibnamefont{Ketterson}},\ }%
  \bibfield{journal}{%
  \bibinfo {journal} {Physical Review B}\ }%
  \textbf{\bibinfo {volume} {80}},\ \bibinfo {pages} {014423} (\bibinfo {year}
  {2009})\BibitemShut{NoStop}%
\bibitem{ShibataandOtani2004}%
  \BibitemOpen
  \bibfield{author}{%
  \bibinfo {author} {\bibfnamefont{J.}~\bibnamefont{Shibata}}\ and\ \bibinfo
  {author} {\bibfnamefont{Y.}~\bibnamefont{Otani}},\ }%
  \bibfield{journal}{%
  \bibinfo {journal} {Physical Review B}\ }%
  \textbf{\bibinfo {volume} {70}},\ \bibinfo {pages} {012404} (\bibinfo {year}
  {2004})\BibitemShut{NoStop}%
\bibitem{Galkinetal2006}%
  \BibitemOpen
  \bibfield{author}{%
  \bibinfo {author} {\bibfnamefont{A.~Y.}\ \bibnamefont{Galkin}}, \bibinfo
  {author} {\bibfnamefont{B.~A.}\ \bibnamefont{Ivanov}},\ and\ \bibinfo
  {author} {\bibfnamefont{C.~E.}\ \bibnamefont{Zaspel}},\ }%
  \bibfield{journal}{%
  \bibinfo {journal} {Physical Review B}\ }%
  \textbf{\bibinfo {volume} {74}},\ \bibinfo {pages} {144419} (\bibinfo {year}
  {2006})\BibitemShut{NoStop}%
\bibitem{Jorzicketal1999}%
  \BibitemOpen
  \bibfield{author}{%
  \bibinfo {author} {\bibfnamefont{J.}~\bibnamefont{Jorzick}}, \bibinfo
  {author} {\bibfnamefont{S.~O.}\ \bibnamefont{Demokritov}}, \bibinfo {author}
  {\bibfnamefont{B.}~\bibnamefont{Hillebrands}}, \bibinfo {author}
  {\bibfnamefont{B.}~\bibnamefont{Bartenlian}}, \bibinfo {author}
  {\bibfnamefont{C.}~\bibnamefont{Chappert}}, \bibinfo {author}
  {\bibfnamefont{D.}~\bibnamefont{Decnanini}}, \bibinfo {author}
  {\bibfnamefont{F.}~\bibnamefont{Rousseaux}},\ and\ \bibinfo {author}
  {\bibfnamefont{E.}~\bibnamefont{Cambril}},\ }%
  \bibfield{journal}{%
  \bibinfo {journal} {Applied Physics Letters}\ }%
  \textbf{\bibinfo {volume} {75}},\ \bibinfo {pages} {3859} (\bibinfo {year}
  {1999})\BibitemShut{NoStop}%
\bibitem{GuslienkoandSlavin2000}%
  \BibitemOpen
  \bibfield{author}{%
  \bibinfo {author} {\bibfnamefont{K.~Y.}\ \bibnamefont{Guslienko}}\ and\
  \bibinfo {author} {\bibfnamefont{A.~N.}\ \bibnamefont{Slavin}},\ }%
  \bibfield{journal}{%
  \bibinfo {journal} {Journal of Applied Physics}\ }%
  \textbf{\bibinfo {volume} {87}},\ \bibinfo {pages} {6337} (\bibinfo {year}
  {2000})\BibitemShut{NoStop}%
\bibitem{Jungetalexp2002}%
  \BibitemOpen
  \bibfield{author}{%
  \bibinfo {author} {\bibfnamefont{S.}~\bibnamefont{Jung}}, \bibinfo {author}
  {\bibfnamefont{B.}~\bibnamefont{Watkins}}, \bibinfo {author}
  {\bibfnamefont{L.}~\bibnamefont{De~Long}}, \bibinfo {author}
  {\bibfnamefont{J.~B.}\ \bibnamefont{Ketterson}},\ and\ \bibinfo {author}
  {\bibfnamefont{V.}~\bibnamefont{Chandrasekhar}},\ }%
  \bibfield{journal}{%
  \bibinfo {journal} {Physical Review B}\ }%
  \textbf{\bibinfo {volume} {66}},\ \bibinfo {pages} {132401} (\bibinfo {year}
  {2002})\BibitemShut{NoStop}%
\bibitem{Jungetaltheory2002}%
  \BibitemOpen
  \bibfield{author}{%
  \bibinfo {author} {\bibfnamefont{S.}~\bibnamefont{Jung}}, \bibinfo {author}
  {\bibfnamefont{J.~B.}\ \bibnamefont{Ketterson}},\ and\ \bibinfo {author}
  {\bibfnamefont{V.}~\bibnamefont{Chandrasekhar}},\ }%
  \bibfield{journal}{%
  \bibinfo {journal} {Physical Review B}\ }%
  \textbf{\bibinfo {volume} {66}},\ \bibinfo {pages} {132405} (\bibinfo {year}
  {2002})\BibitemShut{NoStop}%
\bibitem{PolitiandPini2002}%
  \BibitemOpen
  \bibfield{author}{%
  \bibinfo {author} {\bibfnamefont{P.}~\bibnamefont{Politi}}\ and\ \bibinfo
  {author} {\bibfnamefont{M.}~\bibnamefont{Pini}},\ }%
  \bibfield{journal}{%
  \bibinfo {journal} {Physical Review B}\ }%
  \textbf{\bibinfo {volume} {66}},\ \bibinfo {pages} {214414} (\bibinfo {year}
  {2002})\BibitemShut{NoStop}%
\bibitem{Rivkinetal2004}%
  \BibitemOpen
  \bibfield{author}{%
  \bibinfo {author} {\bibfnamefont{K.}~\bibnamefont{Rivkin}}, \bibinfo {author}
  {\bibfnamefont{A.}~\bibnamefont{Heifetz}}, \bibinfo {author}
  {\bibfnamefont{P.~R.}\ \bibnamefont{Seivert}},\ and\ \bibinfo {author}
  {\bibfnamefont{J.~B.}\ \bibnamefont{Ketterson}},\ }%
  \bibfield{journal}{%
  \bibinfo {journal} {Physical Review B}\ }%
  \textbf{\bibinfo {volume} {70}},\ \bibinfo {pages} {184410} (\bibinfo {year}
  {2004})\BibitemShut{NoStop}%
\bibitem{GubbiottietalJAP2006}%
  \BibitemOpen
  \bibfield{author}{%
  \bibinfo {author} {\bibfnamefont{G.}~\bibnamefont{Gubbiotti}}, \bibinfo
  {author} {\bibfnamefont{M.}~\bibnamefont{Madami}}, \bibinfo {author}
  {\bibfnamefont{S.}~\bibnamefont{Tacchi}}, \bibinfo {author}
  {\bibfnamefont{G.}~\bibnamefont{Carlotti}},\ and\ \bibinfo {author}
  {\bibfnamefont{T.}~\bibnamefont{Okuno}},\ }%
  \bibfield{journal}{%
  \bibinfo {journal} {Journal of Applied Physics}\ }%
  \textbf{\bibinfo {volume} {99}},\ \bibinfo {pages} {08C701} (\bibinfo {year}
  {2006})\BibitemShut{NoStop}%
\bibitem{Giovanninietal2007}%
  \BibitemOpen
  \bibfield{author}{%
  \bibinfo {author} {\bibfnamefont{L.}~\bibnamefont{Giovannini}}, \bibinfo
  {author} {\bibfnamefont{F.}~\bibnamefont{Montoncello}},\ and\ \bibinfo
  {author} {\bibfnamefont{F.}~\bibnamefont{Nizzoli}},\ }%
  \bibfield{journal}{%
  \bibinfo {journal} {Physical Review B}\ }%
  \textbf{\bibinfo {volume} {75}},\ \bibinfo {pages} {024416} (\bibinfo {year}
  {2007})\BibitemShut{NoStop}%
\bibitem{RivkinetalPRB2007}%
  \BibitemOpen
  \bibfield{author}{%
  \bibinfo {author} {\bibfnamefont{K.}~\bibnamefont{Rivkin}}, \bibinfo {author}
  {\bibfnamefont{W.}~\bibnamefont{Saslow}}, \bibinfo {author}
  {\bibfnamefont{L.~E.}\ \bibnamefont{De~Long}},\ and\ \bibinfo {author}
  {\bibfnamefont{J.~B.}\ \bibnamefont{Ketterson}},\ }%
  \bibfield{journal}{%
  \bibinfo {journal} {Physical Review B}\ }%
  \textbf{\bibinfo {volume} {75}},\ \bibinfo {pages} {174408} (\bibinfo {year}
  {2007})\BibitemShut{NoStop}%
\bibitem{RivkinetalJMMM2007}%
  \BibitemOpen
  \bibfield{author}{%
  \bibinfo {author} {\bibfnamefont{K.}~\bibnamefont{Rivkin}}, \bibinfo {author}
  {\bibfnamefont{W.}~\bibnamefont{Xu}}, \bibinfo {author}
  {\bibfnamefont{L.~E.}\ \bibnamefont{De~Long}}, \bibinfo {author}
  {\bibfnamefont{V.~V.}\ \bibnamefont{Metlushko}}, \bibinfo {author}
  {\bibfnamefont{B.}~\bibnamefont{Ilic}},\ and\ \bibinfo {author}
  {\bibfnamefont{J.~B.}\ \bibnamefont{Ketterson}},\ }%
  \bibfield{journal}{%
  \bibinfo {journal} {Journal of Magnetism and Magnetic Materials}\ }%
  \textbf{\bibinfo {volume} {309}},\ \bibinfo {pages} {317} (\bibinfo {year}
  {2007})\BibitemShut{NoStop}%
\bibitem{Kakazeietal2004}%
  \BibitemOpen
  \bibfield{author}{%
  \bibinfo {author} {\bibfnamefont{G.~N.}\ \bibnamefont{Kakazei}}, \bibinfo
  {author} {\bibfnamefont{P.~E.}\ \bibnamefont{Wigen}}, \bibinfo {author}
  {\bibfnamefont{K.~Y.}\ \bibnamefont{Guslienko}}, \bibinfo {author}
  {\bibfnamefont{V.}~\bibnamefont{Novosad}}, \bibinfo {author}
  {\bibfnamefont{A.~N.}\ \bibnamefont{Slavin}}, \bibinfo {author}
  {\bibfnamefont{V.~O.}\ \bibnamefont{Golub}}, \bibinfo {author}
  {\bibfnamefont{N.~A.}\ \bibnamefont{Lesnik}},\ and\ \bibinfo {author}
  {\bibfnamefont{Y.}~\bibnamefont{Otani}},\ }%
  \bibfield{journal}{%
  \bibinfo {journal} {Applied Physics Letters}\ }%
  \textbf{\bibinfo {volume} {85}},\ \bibinfo {pages} {443} (\bibinfo {year}
  {2004})\BibitemShut{NoStop}%
\bibitem{Mewesetal2006}%
  \BibitemOpen
  \bibfield{author}{%
  \bibinfo {author} {\bibfnamefont{T.}~\bibnamefont{Mewes}}, \bibinfo {author}
  {\bibfnamefont{J.}~\bibnamefont{Kim}}, \bibinfo {author}
  {\bibfnamefont{D.~V.}\ \bibnamefont{Pelekov}}, \bibinfo {author}
  {\bibfnamefont{G.~N.}\ \bibnamefont{Kakazei}}, \bibinfo {author}
  {\bibfnamefont{P.~E.}\ \bibnamefont{Wigen}}, \bibinfo {author}
  {\bibfnamefont{S.}~\bibnamefont{Batra}},\ and\ \bibinfo {author}
  {\bibfnamefont{P.~C.}\ \bibnamefont{Hammel}},\ }%
  \bibfield{journal}{%
  \bibinfo {journal} {Physical Review B}\ }%
  \textbf{\bibinfo {volume} {74}},\ \bibinfo {pages} {144424} (\bibinfo {year}
  {2006})\BibitemShut{NoStop}%
\bibitem{Schneideretal2007}%
  \BibitemOpen
  \bibfield{author}{%
  \bibinfo {author} {\bibfnamefont{M.~L.}\ \bibnamefont{Schneider}}, \bibinfo
  {author} {\bibfnamefont{J.~M.}\ \bibnamefont{Shaw}}, \bibinfo {author}
  {\bibfnamefont{A.~B.}\ \bibnamefont{Kos}}, \bibinfo {author}
  {\bibfnamefont{T.}~\bibnamefont{Gerrits}}, \bibinfo {author}
  {\bibfnamefont{T.~J.}\ \bibnamefont{Silva}},\ and\ \bibinfo {author}
  {\bibfnamefont{R.~D.}\ \bibnamefont{McMichael}},\ }%
  \bibfield{journal}{%
  \bibinfo {journal} {Journal of Applied Physics}\ }%
  \textbf{\bibinfo {volume} {102}},\ \bibinfo {pages} {103909} (\bibinfo {year}
  {2007})\BibitemShut{NoStop}%
\bibitem{Rivkinetal2009}%
  \BibitemOpen
  \bibfield{author}{%
  \bibinfo {author} {\bibfnamefont{K.}~\bibnamefont{Rivkin}}, \bibinfo {author}
  {\bibfnamefont{I.~P.}\ \bibnamefont{Nevirkovets}}, \bibinfo {author}
  {\bibfnamefont{O.}~\bibnamefont{Chernyashevskyy}}, \bibinfo {author}
  {\bibfnamefont{J.~B.}\ \bibnamefont{Ketterson}}, \bibinfo {author}
  {\bibfnamefont{B.~K.}\ \bibnamefont{Sarma}},\ and\ \bibinfo {author}
  {\bibfnamefont{V.}~\bibnamefont{Metlushko}},\ }%
  \bibfield{journal}{%
  \bibinfo {journal} {Journal of Magnetism and Magnetic Materials}\ }%
  \textbf{\bibinfo {volume} {321}},\ \bibinfo {pages} {3324} (\bibinfo {year}
  {2009})\BibitemShut{NoStop}%
\bibitem{Shawetal2009}%
  \BibitemOpen
  \bibfield{author}{%
  \bibinfo {author} {\bibfnamefont{J.~M.}\ \bibnamefont{Shaw}}, \bibinfo
  {author} {\bibfnamefont{T.~J.}\ \bibnamefont{Silva}}, \bibinfo {author}
  {\bibfnamefont{M.~L.}\ \bibnamefont{Schneider}},\ and\ \bibinfo {author}
  {\bibfnamefont{R.~D.}\ \bibnamefont{McMichael}},\ }%
  \bibfield{journal}{%
  \bibinfo {journal} {Physical Review B}\ }%
  \textbf{\bibinfo {volume} {79}},\ \bibinfo {pages} {184404} (\bibinfo {year}
  {2009})\BibitemShut{NoStop}%
\bibitem{Nembachetal2009}%
  \BibitemOpen
  \bibfield{author}{%
  \bibinfo {author} {\bibfnamefont{H.~T.}\ \bibnamefont{Nembach}}, \bibinfo
  {author} {\bibfnamefont{H.}~\bibnamefont{Bauer}}, \bibinfo {author}
  {\bibfnamefont{J.~M.}\ \bibnamefont{Shaw}}, \bibinfo {author}
  {\bibfnamefont{M.~L.}\ \bibnamefont{Schneider}},\ and\ \bibinfo {author}
  {\bibfnamefont{T.~J.}\ \bibnamefont{Silva}},\ }%
  \bibfield{journal}{%
  \bibinfo {journal} {Applied Physics Letters}\ }%
  \textbf{\bibinfo {volume} {95}},\ \bibinfo {pages} {062506} (\bibinfo {year}
  {2009})\BibitemShut{NoStop}%
\bibitem{DeckmanandDunsmuir1982}%
  \BibitemOpen
  \bibfield{author}{%
  \bibinfo {author} {\bibfnamefont{H.~W.}\ \bibnamefont{Deckman}}\ and\
  \bibinfo {author} {\bibfnamefont{J.~H.}\ \bibnamefont{Dunsmuir}},\ }%
  \bibfield{journal}{%
  \bibinfo {journal} {Applied Physics Letters}\ }%
  \textbf{\bibinfo {volume} {41}},\ \bibinfo {pages} {377} (\bibinfo {year}
  {1982})\BibitemShut{NoStop}%
\bibitem{HulteenandVanDuyne1995}%
  \BibitemOpen
  \bibfield{author}{%
  \bibinfo {author} {\bibfnamefont{J.~C.}\ \bibnamefont{Hulteen}}\ and\
  \bibinfo {author} {\bibfnamefont{R.~P.}\ \bibnamefont{Van~Duyne}},\ }%
  \bibfield{journal}{%
  \bibinfo {journal} {Journal of Vacuum Science and Technology}\ }%
  \textbf{\bibinfo {volume} {13}},\ \bibinfo {pages} {1553} (\bibinfo {year}
  {1995})\BibitemShut{NoStop}%
\bibitem{Hulteenetal1999}%
  \BibitemOpen
  \bibfield{author}{%
  \bibinfo {author} {\bibfnamefont{J.~C.}\ \bibnamefont{Hulteen}}, \bibinfo
  {author} {\bibfnamefont{D.~A.}\ \bibnamefont{Trichel}}, \bibinfo {author}
  {\bibfnamefont{M.~T.}\ \bibnamefont{Smith}}, \bibinfo {author}
  {\bibfnamefont{M.~L.}\ \bibnamefont{Duval}}, \bibinfo {author}
  {\bibfnamefont{T.~R.}\ \bibnamefont{Jensen}},\ and\ \bibinfo {author}
  {\bibfnamefont{R.~P.}\ \bibnamefont{Van~Duyne}},\ }%
  \bibfield{journal}{%
  \bibinfo {journal} {Journal of Physical Chemistry}\ }%
  \textbf{\bibinfo {volume} {103}},\ \bibinfo {pages} {2394} (\bibinfo {year}
  {1999})\BibitemShut{NoStop}%
\bibitem{Jensenetal1999}%
  \BibitemOpen
  \bibfield{author}{%
  \bibinfo {author} {\bibfnamefont{T.~R.}\ \bibnamefont{Jensen}}, \bibinfo
  {author} {\bibfnamefont{G.~C.}\ \bibnamefont{Schatz}},\ and\ \bibinfo
  {author} {\bibfnamefont{R.~P.}\ \bibnamefont{Van~Duyne}},\ }%
  \bibfield{journal}{%
  \bibinfo {journal} {Journal of Physical Chemistry}\ }%
  \textbf{\bibinfo {volume} {103}},\ \bibinfo {pages} {2394} (\bibinfo {year}
  {1999})\BibitemShut{NoStop}%
\bibitem{Haginoyaetal1997}%
  \BibitemOpen
  \bibfield{author}{%
  \bibinfo {author} {\bibfnamefont{C.}~\bibnamefont{Haginoya}}, \bibinfo
  {author} {\bibfnamefont{M.}~\bibnamefont{Ishibashi}},\ and\ \bibinfo {author}
  {\bibfnamefont{K.}~\bibnamefont{Koike}},\ }%
  \bibfield{journal}{%
  \bibinfo {journal} {Applied Physics Letters}\ }%
  \textbf{\bibinfo {volume} {71}},\ \bibinfo {pages} {2934} (\bibinfo {year}
  {1997})\BibitemShut{NoStop}%
\bibitem{Burmeisteretal1997}%
  \BibitemOpen
  \bibfield{author}{%
  \bibinfo {author} {\bibfnamefont{F.}~\bibnamefont{Burmeister}}, \bibinfo
  {author} {\bibfnamefont{C.}~\bibnamefont{Sch{\"a}fle}}, \bibinfo {author}
  {\bibfnamefont{T.}~\bibnamefont{Matthes}}, \bibinfo {author}
  {\bibfnamefont{M.}~\bibnamefont{B{\"o}hmisch}}, \bibinfo {author}
  {\bibfnamefont{J.}~\bibnamefont{Boneberg}},\ and\ \bibinfo {author}
  {\bibfnamefont{P.}~\bibnamefont{Leiderer}},\ }%
  \bibfield{journal}{%
  \bibinfo {journal} {Langmuir}\ }%
  \textbf{\bibinfo {volume} {13}},\ \bibinfo {pages} {2983} (\bibinfo {year}
  {1997})\BibitemShut{NoStop}%
\bibitem{Ormondeetal2004}%
  \BibitemOpen
  \bibfield{author}{%
  \bibinfo {author} {\bibfnamefont{A.~D.}\ \bibnamefont{Ormonde}}, \bibinfo
  {author} {\bibfnamefont{E.~C.~M.}\ \bibnamefont{Hicks}}, \bibinfo {author}
  {\bibfnamefont{J.}~\bibnamefont{Castillo}},\ and\ \bibinfo {author}
  {\bibfnamefont{R.~P.}\ \bibnamefont{Van~Duyne}},\ }%
  \bibfield{journal}{%
  \bibinfo {journal} {Langmuir}\ }%
  \textbf{\bibinfo {volume} {20}},\ \bibinfo {pages} {6927} (\bibinfo {year}
  {2004})\BibitemShut{NoStop}%
\bibitem{Rybczynskietal2003}%
  \BibitemOpen
  \bibfield{author}{%
  \bibinfo {author} {\bibfnamefont{J.}~\bibnamefont{Rybczynski}}, \bibinfo
  {author} {\bibfnamefont{U.}~\bibnamefont{Ebels}},\ and\ \bibinfo {author}
  {\bibfnamefont{M.}~\bibnamefont{Giersig}},\ }%
  \bibfield{journal}{%
  \bibinfo {journal} {Colloids and Surfaces A}\ }%
  \textbf{\bibinfo {volume} {219}},\ \bibinfo {pages} {1} (\bibinfo {year}
  {2003})\BibitemShut{NoStop}%
\bibitem{Weekesetal2004}%
  \BibitemOpen
  \bibfield{author}{%
  \bibinfo {author} {\bibfnamefont{S.~M.}\ \bibnamefont{Weekes}}, \bibinfo
  {author} {\bibfnamefont{F.~Y.}\ \bibnamefont{Ogrin}},\ and\ \bibinfo {author}
  {\bibfnamefont{W.~A.}\ \bibnamefont{Murray}},\ }%
  \bibfield{journal}{%
  \bibinfo {journal} {Langmuir}\ }%
  \textbf{\bibinfo {volume} {20}},\ \bibinfo {pages} {11208} (\bibinfo {year}
  {2004})\BibitemShut{NoStop}%
\bibitem{Weekesetal2007}%
  \BibitemOpen
  \bibfield{author}{%
  \bibinfo {author} {\bibfnamefont{S.~M.}\ \bibnamefont{Weekes}}, \bibinfo
  {author} {\bibfnamefont{F.~Y.}\ \bibnamefont{Ogrin}}, \bibinfo {author}
  {\bibfnamefont{W.~A.}\ \bibnamefont{Murray}},\ and\ \bibinfo {author}
  {\bibfnamefont{P.~S.}\ \bibnamefont{Keatley}},\ }%
  \bibfield{journal}{%
  \bibinfo {journal} {Langmuir}\ }%
  \textbf{\bibinfo {volume} {23}},\ \bibinfo {pages} {1057} (\bibinfo {year}
  {2007})\BibitemShut{NoStop}%
\bibitem{Lietal2008}%
  \BibitemOpen
  \bibfield{author}{%
  \bibinfo {author} {\bibfnamefont{H.}~\bibnamefont{Li}}, \bibinfo {author}
  {\bibfnamefont{J.}~\bibnamefont{Low}}, \bibinfo {author}
  {\bibfnamefont{K.~S.}\ \bibnamefont{Brown}},\ and\ \bibinfo {author}
  {\bibfnamefont{N.}~\bibnamefont{Wu}},\ }%
  \bibfield{journal}{%
  \bibinfo {journal} {IEEE Sensors Journal}\ }%
  \textbf{\bibinfo {volume} {8}},\ \bibinfo {pages} {880} (\bibinfo {year}
  {2008})\BibitemShut{NoStop}%
\bibitem{Zhangetal2008}%
  \BibitemOpen
  \bibfield{author}{%
  \bibinfo {author} {\bibfnamefont{Y.}~\bibnamefont{Zhang}}, \bibinfo {author}
  {\bibfnamefont{X.}~\bibnamefont{Wang}}, \bibinfo {author}
  {\bibfnamefont{Y.}~\bibnamefont{Wang}}, \bibinfo {author}
  {\bibfnamefont{H.}~\bibnamefont{Liu}},\ and\ \bibinfo {author}
  {\bibfnamefont{J.}~\bibnamefont{Yang}},\ }%
  \bibfield{journal}{%
  \bibinfo {journal} {Journal of Alloys and Compounds}\ }%
  \textbf{\bibinfo {volume} {452}},\ \bibinfo {pages} {473} (\bibinfo {year}
  {2008})\BibitemShut{NoStop}%
\bibitem{Couniletal2004}%
  \BibitemOpen
  \bibfield{author}{%
  \bibinfo {author} {\bibfnamefont{G.}~\bibnamefont{Counil}}, \bibinfo {author}
  {\bibfnamefont{P.}~\bibnamefont{Crozat}}, \bibinfo {author}
  {\bibfnamefont{T.}~\bibnamefont{Devolder}}, \bibinfo {author}
  {\bibfnamefont{C.}~\bibnamefont{Chappert}}, \bibinfo {author}
  {\bibfnamefont{S.}~\bibnamefont{Zoll}},\ and\ \bibinfo {author}
  {\bibfnamefont{R.}~\bibnamefont{Fournel}},\ }%
  \bibfield{journal}{%
  \bibinfo {journal} {IEEE Transactions on Magnetics}\ }%
  \textbf{\bibinfo {volume} {42}},\ \bibinfo {pages} {3321} (\bibinfo {year}
  {2004})\BibitemShut{NoStop}%
\bibitem{Kittel1996}%
  \BibitemOpen
  \bibfield{author}{%
  \bibinfo {author} {\bibfnamefont{C.}~\bibnamefont{Kittel}},\ }%
  \emph{\bibinfo {title} {Introduction to solid state physics}},\ \bibinfo
  {edition} {seventh}\ ed.\ (\bibinfo {publisher} {John Wiley and Sons, Inc.},\
  \bibinfo {year} {1996})\BibitemShut{NoStop}%
\bibitem{OOMMFweb}%
  \BibitemOpen
  \bibfield{author}{%
  \bibinfo {author} {\bibfnamefont{M.~J.}\ \bibnamefont{Donahue}}\ and\
  \bibinfo {author} {\bibfnamefont{D.~G.}\ \bibnamefont{Porter}},\ }%
  \enquote{\bibinfo {title} {{Object Oriented Micro-Magnetic Framework
  (OOMMF)}},}\ \bibinfo {note}
  {\texttt{http://math.nist.gov/oommf/}}\BibitemShut{NoStop}%
\bibitem{Kostylev2009}%
  \BibitemOpen
  \bibfield{author}{%
  \bibinfo {author} {\bibfnamefont{M.}~\bibnamefont{Kostylev}},\ }%
  \bibfield{journal}{%
  \bibinfo {journal} {Journal of Applied Physics}\ }%
  \textbf{\bibinfo {volume} {106}},\ \bibinfo {pages} {043903} (\bibinfo {year}
  {2009})\BibitemShut{NoStop}%
\bibitem{Kalarickaletal2006}%
  \BibitemOpen
  \bibfield{author}{%
  \bibinfo {author} {\bibfnamefont{S.~S.}\ \bibnamefont{Kalarickal}}, \bibinfo
  {author} {\bibfnamefont{P.}~\bibnamefont{Krivosik}}, \bibinfo {author}
  {\bibfnamefont{M.}~\bibnamefont{Wu}}, \bibinfo {author}
  {\bibfnamefont{C.~E.}\ \bibnamefont{Patton}}, \bibinfo {author}
  {\bibfnamefont{M.~L.}\ \bibnamefont{Schneider}}, \bibinfo {author}
  {\bibfnamefont{P.}~\bibnamefont{Kabos}}, \bibinfo {author}
  {\bibfnamefont{T.~J.}\ \bibnamefont{Silva}},\ and\ \bibinfo {author}
  {\bibfnamefont{J.~P.}\ \bibnamefont{Nibarger}},\ }%
  \bibfield{journal}{%
  \bibinfo {journal} {Journal of Applied Physics}\ }%
  \textbf{\bibinfo {volume} {99}},\ \bibinfo {pages} {093909} (\bibinfo {year}
  {2006})\BibitemShut{NoStop}%
\bibitem{LandauandLifshitz1935}%
  \BibitemOpen
  \bibfield{author}{%
  \bibinfo {author} {\bibfnamefont{L.}~\bibnamefont{Landau}}\ and\ \bibinfo
  {author} {\bibfnamefont{E.}~\bibnamefont{Lifshitz}},\ }%
  \bibfield{journal}{%
  \bibinfo {journal} {Physikalische Zeitschrift der Sowjetunion}\ }%
  \textbf{\bibinfo {volume} {8}},\ \bibinfo {pages} {153} (\bibinfo {year}
  {1935})\BibitemShut{NoStop}%
\bibitem{GilbertPhDThesis}%
  \BibitemOpen
  \bibfield{author}{%
  \bibinfo {author} {\bibfnamefont{T.~L.}\ \bibnamefont{Gilbert}},\ }%
  Ph.D. thesis,\ \bibinfo {school} {Illinois Institute of Technology} (\bibinfo
  {year} {1956})\BibitemShut{NoStop}%
\bibitem{Lenzetal2006}%
  \BibitemOpen
  \bibfield{author}{%
  \bibinfo {author} {\bibfnamefont{K.}~\bibnamefont{Lenz}}, \bibinfo {author}
  {\bibfnamefont{H.}~\bibnamefont{Wende}}, \bibinfo {author}
  {\bibfnamefont{W.}~\bibnamefont{Kuch}}, \bibinfo {author}
  {\bibfnamefont{K.}~\bibnamefont{Baberschke}}, \bibinfo {author}
  {\bibfnamefont{K.}~\bibnamefont{Nagy}},\ and\ \bibinfo {author}
  {\bibfnamefont{A.}~\bibnamefont{J{\'a}nossy}},\ }%
  \bibfield{journal}{%
  \bibinfo {journal} {Physical Review B}\ }%
  \textbf{\bibinfo {volume} {73}},\ \bibinfo {pages} {144424} (\bibinfo {year}
  {2006})\BibitemShut{NoStop}%
\bibitem{Bloch1946}%
  \BibitemOpen
  \bibfield{author}{%
  \bibinfo {author} {\bibfnamefont{F.}~\bibnamefont{Bloch}},\ }%
  \bibfield{journal}{%
  \bibinfo {journal} {Physical Review}\ }%
  \textbf{\bibinfo {volume} {70}},\ \bibinfo {pages} {460} (\bibinfo {year}
  {1946})\BibitemShut{NoStop}%
\bibitem{Sankeyetal2006}%
  \BibitemOpen
  \bibfield{author}{%
  \bibinfo {author} {\bibfnamefont{J.~C.}\ \bibnamefont{Sankey}}, \bibinfo
  {author} {\bibfnamefont{P.~M.}\ \bibnamefont{Braganca}}, \bibinfo {author}
  {\bibfnamefont{A.~G.~F.}\ \bibnamefont{Garcia}}, \bibinfo {author}
  {\bibfnamefont{I.~N.}\ \bibnamefont{Krivorotov}}, \bibinfo {author}
  {\bibfnamefont{R.~A.}\ \bibnamefont{Buhrman}},\ and\ \bibinfo {author}
  {\bibfnamefont{D.~C.}\ \bibnamefont{Ralph}},\ }%
  \bibfield{journal}{%
  \bibinfo {journal} {Physical Review Letters}\ }%
  \textbf{\bibinfo {volume} {96}},\ \bibinfo {pages} {227601} (\bibinfo {year}
  {2006})\BibitemShut{NoStop}%
\bibitem{Gubbiottietal2003}%
  \BibitemOpen
  \bibfield{author}{%
  \bibinfo {author} {\bibfnamefont{G.}~\bibnamefont{Gubbiotti}}, \bibinfo
  {author} {\bibfnamefont{G.}~\bibnamefont{Carlotti}}, \bibinfo {author}
  {\bibfnamefont{R.}~\bibnamefont{Ziveri}}, \bibinfo {author}
  {\bibfnamefont{F.}~\bibnamefont{Nizzoli}}, \bibinfo {author}
  {\bibfnamefont{T.}~\bibnamefont{Okuno}},\ and\ \bibinfo {author}
  {\bibfnamefont{T.}~\bibnamefont{Shinjo}},\ }%
  \bibfield{journal}{%
  \bibinfo {journal} {Journal of Applied Physics}\ }%
  \textbf{\bibinfo {volume} {93}},\ \bibinfo {pages} {7607} (\bibinfo {year}
  {2003})\BibitemShut{NoStop}%
\bibitem{AriasandMills1999}%
  \BibitemOpen
  \bibfield{author}{%
  \bibinfo {author} {\bibfnamefont{R.}~\bibnamefont{Arias}}\ and\ \bibinfo
  {author} {\bibfnamefont{D.~L.}\ \bibnamefont{Mills}},\ }%
  \bibfield{journal}{%
  \bibinfo {journal} {Physical Review B}\ }%
  \textbf{\bibinfo {volume} {60}},\ \bibinfo {pages} {7395} (\bibinfo {year}
  {1999})\BibitemShut{NoStop}%
\end{thebibliography}
\end{document}